\documentclass[12pt]{article}

\usepackage{graphicx}
\usepackage{palatino}

\usepackage{geometry}
\usepackage{amsmath,amssymb,amstext} 
\usepackage{hyperref}
\usepackage{subfigure}
\usepackage{color}
\usepackage{multirow} 
\usepackage{rotating} 
\usepackage{longtable} 
\usepackage{array} 
\usepackage{natbib}
\usepackage{float}
\usepackage{fmtcount}

\renewcommand{\large}{\fontsize{12}{15}\selectfont}
\renewcommand{\Large}{\fontsize{14}{18}\selectfont}

\renewcommand{\Huge}{\fontsize{24}{30}\selectfont}
\makeatletter
\def\nrfootnote{\@ifnextchar[\@xfootnote{\stepcounter\@mpfn
\protected@xdef\@thefnmark{\thempfn}%
\@footnotetext}}

\def\blfootnote{\xdef\@thefnmark{}\@footnotetext}
\makeatother

\begin{document}
\renewcommand{\arraystretch}{0.66666666667}

\begin{titlepage}

\fontfamily{ppl}\selectfont

\hspace*{-1cm}
\hbox{
\includegraphics[width=4cm]{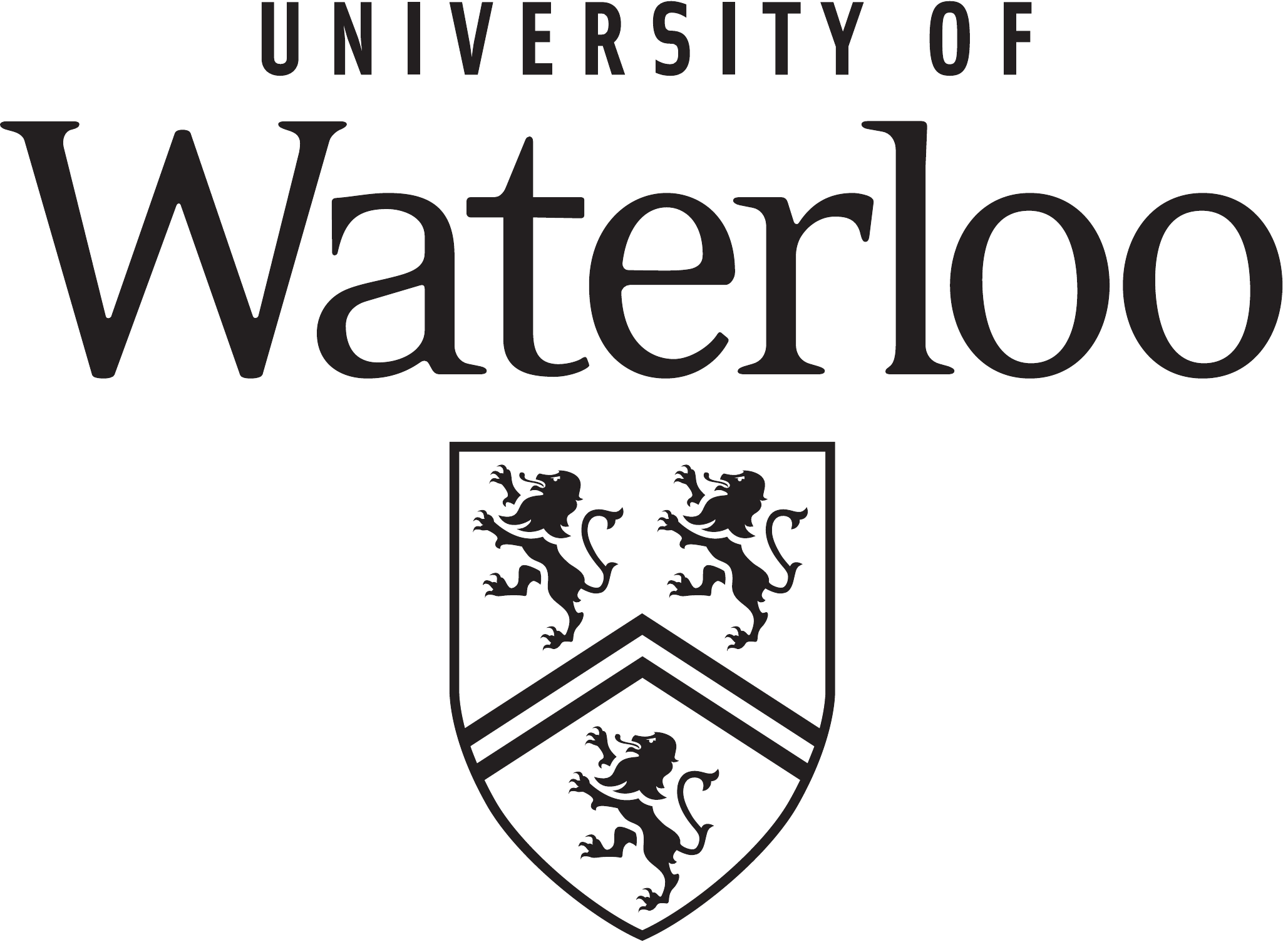}
\hspace{1cm}
\parbox[b][2cm][t]{12cm}{
\noindent
\fontfamily{phv}\selectfont
\fontsize{8}{10}\selectfont
{\bfseries
DEPARTMENT OF STATISTICS AND ACTUARIAL SCIENCE} \\
University of Waterloo,
200 University Avenue West
Waterloo, Ontario, Canada, N2L 3G1 \\
519-888-4567
\hspace{1mm}$|$\hspace{1mm}
Fax:  519-746-1875
\hspace{1mm}$|$\hspace{1mm}
\bfseries www.stats.uwaterloo.ca
}}
\vspace{2cm}

\begin{center}
{\bfseries

\Large
UNIVERSITY OF WATERLOO\\

\vspace{0.1in}

\Large
DEPARTMENT OF STATISTICS AND ACTUARIAL SCIENCE

\vspace{0.5in}

\large
WORKING PAPER

\vspace{0.1in}
}
\vspace{0.5in}

{\Huge \bfseries Assessing the Health of Richibucto Estuary with the Latent Health Factor Index}\footnote{On 2013-05-01, a revised version of this article was accepted for publication in {\it PLoS One}. DOI: 10.1371/journal.pone.0065697}

\vspace{0.5in}

{\ {\bf Margaret Wu} \\
Business Methods Survey Division, Statistics Canada
}

\vspace{0.2in}

{\ {\bf Grace S.~Chiu}\footnote{Corresponding author. E-mail: gchiu.off.campus@gmail.com} \\
University of Waterloo, Dept.~of Statistics \& Actuarial Science\footnote{Until September 2012, prior to which the current version of this article was written.} \\
}

\vspace{0.2in}

{\ {\bf Lin Lu} \\
McGregor Geoscience Ltd.
}


\end{center}

\end{titlepage}


\newcommand{\ul}{\underline}
\newcommand{\be}{\begin{enumerate}}
\newcommand{\ee}{\end{enumerate}}
\newcommand{\bi}{\begin{itemize}}
\newcommand{\ei}{\end{itemize}}
\newcommand{\bc}{\begin{center}}
\newcommand{\ec}{\end{center}}
\newcommand{\beq}{\begin{equation}}
\newcommand{\eeq}{\end{equation}}
\newcommand{\nn}{\nonumber}
\newcommand{\bs}{\boldsymbol}
\newcommand{\wh}{\widehat}
\newcommand{\eps}{\varepsilon}
\newcommand{\avg}{\overline}
\newcommand{\stck}{\stackrel}
\newcommand{\iid}{\stck{\text{iid}}{\sim}}
\newcommand{\ind}{\stck{\text{ind}}{\sim}}

\subsection*{Abstract}
\small
The ability to quantitatively assess the health of an ecosystem is often of great interest to those tasked with monitoring and conserving ecosystems. For decades, research in this area has relied upon multimetric indices of various forms. Although indices may be numbers, many are constructed based on procedures that are highly qualitative in nature, thus limiting the quantitative rigour of the practical interpretations made from these indices. The statistical modelling approach to construct the latent health factor index (LHFI) was recently developed to express ecological data, collected to construct conventional multimetric health indices, in a rigorous quantitative model that integrates qualitative features of ecosystem health and preconceived ecological relationships among such features. This hierarchical modelling approach allows (a) statistical inference of health for observed sites and (b) prediction of health for unobserved sites, all accompanied by formal uncertainty statements. Thus far, the LHFI approach has been demonstrated and validated on freshwater ecosystems. The goal of this paper is to adapt this approach to modelling estuarine ecosystem health, particularly that of the previously unassessed system in Richibucto in New Brunswick, Canada. Field data correspond to biotic health metrics that constitute the AZTI marine biotic index (AMBI) and abiotic predictors preconceived to influence biota. We also briefly discuss related LHFI research involving additional metrics that form the infaunal trophic index (ITI). Our paper is the first to construct a scientifically sensible model to rigorously identify the collective explanatory capacity of salinity, distance downstream, channel depth, and silt-clay content --- all regarded {\it a priori} as qualitatively important abiotic drivers --- towards site health in the Richibucto ecosystem.
\par\noindent
{\it Keywords:} AMBI; Bayesian statistics; hierarchical modelling; infaunal trophic index; Markov chain Monte Carlo; statistical inference
\normalsize

\section{Existing Methods to Quantify Ecosystem Health}

Assessment of the ``health'' of an ecosystem is often of great importance to those interested in the monitoring and conservation of ecosystems. Health is a complex concept often involving many diverse factors, and therefore is not straightforward to quantify. A popular method to estimate the health of an ecosystem is through one or more multimetric indices, each of which is a scalar number collapsed from several indicator variables of health, or metrics. Ecosystem health metrics are frequently measures of faunal abundance and diversity. For aquatic ecosystems, these biotic metrics often focus on benthic populations --- organisms living on or in the sediment at the bottom of a body of water --- since they are useful indicators of underlying health conditions \citep{bilyard, dauer}. For example, the AZTI\footnote{Marine and Food Technological Centre (\url{http://www.azti.es}).} marine biotic index (AMBI) \citep{borja} is a quantitative measure of health for an estuarine ecosystem based on the sample counts of categorised benthos. Its popularity is evident from its use across the globe, including Africa \citep{bazairi}, Asia \citep{cai}, Europe \citep{medeiros}, North America \citep{teixeira}, and South America \citep{muniz}. 

AMBI and other common multimetric indices, e.g.~infaunal trophic index (ITI) \citep{word}, estuarine biotic integrity index \citep{deegan}, benthic response index \citep{smith}, benthic quality index \citep{rosenberg}, infaunal quality index \citep{mackie}, have the main appeal that they are conceptually simple and thus easily interpretable. They also contain a high amount of biological content from subject-matter scientists being involved at all stages of the design of the index. On the other hand, many of these stages during index construction can involve a non-trivial amount of arbitrariness. Consequently, rigorous evaluation of index reliability and other quantitative aspects is difficult with conventional indices: for example, detecting relationships between health and environmental or impact-related covariates such as water depth or urbanisation; and formally assessing the uncertainty in these estimates of health. Recent multi-step approaches towards addressing such concerns \citep[e.g.][]{smith,johnston} do not address propagation of uncertainty from one step to another, thereby resulting in inference that is less reliable than that from an integrated statistical methodology. \citet{chiu.guttorp} proposed the SHIPSL approach, a statistically enhanced multimetric index construction scheme that improves various quantitative aspects of conventional indices \citep{dobbie}, although it and others share unresolved issues such as non-transferability in space or time, and the need for follow-up analyses to determine its relationship with non-faunal (abiotic) variables in method evaluation or policy-making contexts.

Recently \citet{chiu.paper} devised the {\em latent health factor index} (LHFI), a novel statistical model-based ecological index aimed to retain the advantages of conventional multimetric indices while addressing some of their shortcomings. The LHFI methodology involves a multi-level analysis of covariance generalised linear mixed-effects (regression) model \citep[e.g.][]{textbook}, or ANOCOVA GLMM: instead of being treated as measures of health, metrics are regarded as {\em indicators} of underlying health conditions; thus, these indicators are regressed as response variables upon a latent health quantity (latent since it is not directly observable) which is site-specific, forming the main level of the regression; health in turn can be regressed upon available covariates, such as environmental (e.g.~salinity, silt-clay content) and impact related (e.g.~urbanisation) variables, forming the optional sublevel in the model hierarchy.

With data on metrics and covariates, latent health can be estimated as a scalar,
so that interpretability is retained; the estimated quantity is the value of the index. Additionally, the effect of the covariates on health can be estimated in a single integrated statistical framework. Importantly, statistical modelling is what directly produces the health index under the integrated LHFI framework, as opposed to being employed merely to select relevant metrics before index construction \citep[e.g.~in][]{deegan} or to evaluate the resulting index \citep[e.g.~in][]{borja2}. Thus, the LHFI is much more rigorous than conventional indices, as its definition utilises universal modelling practices for the definition of the index; its hierarchical modelling framework also allows for comprehensive statistical inference without the need for sequential analyses through which the propagation of uncertainty is lost from one analysis to the next. As well, the LHFI model, once specified for a set of existing sites, allows for cost effective yet rigorous interpolation of health for a new site: prediction can be accomplished simply with information on covariates at this new site, thus bypassing the expensive benthic taxonomic laboratory procedures that are required to gather the metric data as required by conventional indices. These desirable properties are gained without sacrificing biological integrity which can be embedded through subject-matter expertise in the identification of useful metrics and covariates for constructing the LHFI. It is also straightforward to use the LHFI framework to handle data that have certain spatial and/or temporal features, thus resolving the non-transferability issue of previous indices.

Recently, \citet{schliep} integrated formal point-referenced spatial modelling \cite{banerjee} with LHFI principles to model the hierarchical relationship among four levels of quantities: five-point ordinal health metrics (scaled from ``poor'' to ``excellent''), latent continuous quantities that determine the ordinal metrics, latent health, and drivers of health that pertain to geographical and environmental characteristics. They illustrate the type of unified statistical inference that can be drawn from such an LHFI-based approach for assessing biotic integrity of river basins in Colorado, USA. In contrast, \citet{chiu.paper} and \citet{wu} directly model the quantitative health indicators based upon which ordinal metrics such as those of \citet{schliep} are defined. This avoids loss of information due to the mapping of quantitative health metrics to a coarse ordinal scale.

More information on popular indices, the LHFI, and their properties are given by \citet{chiu.paper} and \citet{wu}.

\section{Previous LHFIs for the Richibucto Estuary}

\begin{figure}[t]
\centering
\includegraphics[width=5in]{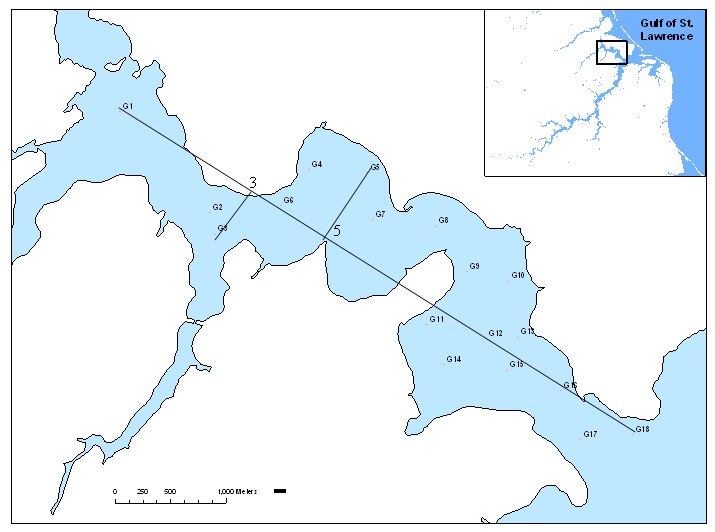}
\caption{Map of Richibucto estuary with the 18 monitored sites (labelled `G' followed by the site number); the straight lines illustrate the calculation of {\em distance downstream} (DD) for Sites 3 and 5.\label{map}}
\end{figure}

The LHFI modelling methodology was first demonstrated and validated on freshwater ecosystems by \citet{chiu.paper}. Following this, \citet{wu} applied the methodology successfully to an estuarine ecosystem, utilising the dataset from \citet{luke} on the previously unassessed Richibucto estuary in the Canadian province of New Brunswick. The data were collected in the estuary at 18 sites (Fig.~\ref{map}). Sites 2, 4--7, and 14 were closest to active oyster farms. Oyster farming activity is perceived to impact site health through its direct influence on sediment properties, although \cite{luke} report that different biotic indicators show different types of association with proximity to oyster farms. For example, macrobenthic faunal abundance is medium for 5 of these 6 sites but high for Site 2; Shannon's diversity is relatively even among all 18 sites with a slight upward trend as the site moves away from the upper channel instead of from an oyster farm. This lack of obvious association remains despite these authors' efforts to consider certain individual species as separate health indicators. This motivated us to develop LHFI models for Richibucto based on indicator metrics (Table \ref{table-metrics}) used to construct the AMBI and ITI, two popular estuarine ecosystem health indices. Specifically, AMBI and ITI metrics are better tailored to estuarine ecosystems than the generic indicators of abundance, richness, and diversity; and they are more comprehensive than indicators based on single species. However, biotic health indicators alone do not explicitly reveal the collective impact on overall health from abiotic variables, including sediment properties (which, in this case, can be affected by oyster farming activity), oceanographic properties (salinity, water temperature, etc.), and their interactions. The work by \cite{wu} is the precursor to our current paper.

\citet{wu} developed two sets of LHFI models, the first using only metrics from AMBI (denoted by LHFI-A), and the second using metrics from both AMBI and ITI (denoted by LHFI-A-I). These models were considered in a Bayesian statistical framework and implemented using Markov chain Monte Carlo (MCMC) techniques. Both sets of models were successful in that they were able to make reasonable distinctions between health levels at different sites, while allowing rigorous assessment of reliability. As well, the resulting health estimates were justifiable by subject-matter expertise.
\begin{table}[t]
\caption{Metrics, based on definitions of the AMBI and ITI, that are used here and by \citet{wu} to construct LHFIs for the Richibucto estuary. \label{table-metrics}}
\centering\small
\begin{tabular}{clc}\\ \hline
&& Preconceived \\
&& association \\
& {\bf AMBI abundance}$^\text{\ref{abund-note}}$ {\bf metric} & w/ health\\ \hline \\
1 & species (including specialist carnivores and some deposit-feeding & + \\
& tubicolous polychaetes) very sensitive to organic enrichment and \\
& present under unpolluted conditions \\\\
2 & species (including suspension feeders, less selective carnivores and & $\pm^\text{\ref{pm-note}}$ \\
& scavengers) indifferent to enrichment, always present in low densities \\
& with non-significant variations with time \\\\
3 & species tolerant to excess organic matter enrichment (including & $-$ \\
& surface deposit-feeding species, e.g.~tubicolous spionids) \\\\
4 & second-order oppotunistic species; mainly small-sized polychaetes: & $-$ \\
& subsurface deposit-feeders, e.g.~cirratulids \\\\
5 & first-order opportunistic species: deposit-feeders, which proliferate & $-$ \\
& in reduced sediments \\ \hline \\
& {\bf ITI abundance metric}$^\text{\ref{iti-note}}$ \\ \hline \\
1 & suspension feeders: feed on detritus from the water column and & + \\
& usually lack sediment grains in their stomach contents \\\\
2 & interface / surface detrital feeders: obtain the same types of food as & + \\
& suspension feeders but usually from the upper 0.5 cm of the sediment \\\\
3 & deposit feeders: invertebrates (including carnivores); generally feed & $\pm$ \\
& from the top few cm of the sediment and feed on encrusted mineral \\
& aggregates, deposit particles or biological remains \\\\
4 & specialised environment feeders: mobile burrowers that feed on & $-$ \\
& deposited organic material; all adapted to live in highly anaerobic \\
& sediment \\ \hline \\
\end{tabular}
\end{table}
\nrfootnote{The fraction of organisms with the specified characteristics out of all benthic organisms in the grab sample.\label{abund-note}}%
\nrfootnote{Neither clearly positive nor clearly negative.\label{pm-note}}%
\nrfootnote{As described by \citet{cromey}.\label{iti-note}}

\subsection{What May Influence Richibucto's Health?}\label{sec-influence}
AMBI metrics, which for the most part pertain to organic enrichment, were the main focus when \citet{wu} constructed the Richibucto LHFIs. This was because prior to Wu's analyses, it had been perceived that benthic fauna in Richibucto were related to organic enrichment, as well as freshwater input (salinity gradient), variability of particle size and topography (channel and water depth) \citep{luke}. These non-enrichment aspects of the estuary were observed as covariates alongside benthic fauna; see \citet{luke} for details.

As such, in addition to assessing the health of the Richibucto system, the above LHFI models were used to 
investigate which and how covariates may influence health as reflected by biotic metrics. As discussed by \citet{chiu.paper} and \citet{wu}, a thorough understanding of the relationship between covariates and health is key to rigorous yet cost effective interpolation of site health, and could prove to be an enormous asset to scientists. To this end, each of the above \mbox{LHFI-A} and LHFI-A-I models had been implemented with a different combination of covariates.
Several covariates and interactions were found to have a statistically significant relationship with health.\footnote{Significance here refers to the regression coefficient having an interval with a reasonably high Bayesian credible level (e.g.~0.8) while excluding the value 0.} For LHFI-A, a model with site-associated covariates \textit{log(SC)}, \textit{log(depth)}, \textit{salinity}, and interaction \textit{log(SC)$\times$log(depth)}, and another model with a single covariate \textit{distance downstream} (DD), were the two best-fitting models among those investigated. For LHFI-A-I, there were three best models: two corresponded to the same sets of covariates as those for LHFI-A, and another model with covariates log(depth), log(SC) and their interaction. SC denotes the fraction of silt-clay (grains of size $<$63 $\mu$m) out of the sediment pooled from all (2 to 3) grab sample replicates. Depth is the distance (m) from the water surface to the estuary bed at the location of the site from which grab samples were obtained. Salinity (parts per thousand) was measured based on one {\em in situ} water sample obtained at the site. To determine DD (km), first a straight line was extended from the western-most site (Site 1) to the eastern-most site (Site 18); DD of any site is defined as the distance between the site's perpendicular projection onto the straight line and Site 1 (Fig.~\ref{map}). Other site-specific covariates also considered but found to be insignificant or confounded with others were {\em water temperature} ($^\text{o}$C), {\em time of year} (September or October), {\em median grain size} of sediment, {\em sorting} (a unitless measure of variability of grain size) and {\em organic content} (\%).
\begin{table}[t]
\centering
\caption{Sample correlation coefficients among covariates for latent health of Richibucto sites.\label{tab.cor}}
\begin{tabular}{c|cccc}
           & DD & salinity & log(depth) & log(SC) \\ \hline
DD         & 1 & 0.88 & 0.16 & $-$0.47 \\
salinity   &   & 1    & 0.23 & $-$0.33 \\
log(depth) &   &      & 1    & $-$0.41 \\
log(SC)    &   &      &      & 1
\end{tabular}
\end{table}

However, attempts by \citet{wu} to include covariates from various best-fitting models together in a single LHFI-A or LHFI-A-I model were unsatisfactory; in such combined models, DD remained highly significant, while all other covariates and their interactions were no longer significant at a reasonable credible level. Indeed, salinity and DD are highly correlated (Table \ref{tab.cor}), and it was unsurprising that the two are not simultaneously significant. However, no strong correlation exist among log(depth), log(SC), and DD (Table \ref{tab.cor}), and so why did DD ``trump'' all others in a combined model, despite non-DD covariates being significant when DD was absent? This question had yet to be addressed. As well, while relationships between health and covariates were quite strong for the LHFI-A models, they were less clear for LHFI-A-I models (significance at credible levels $\approx$60--85\% in the best-fitting LHFI-A-I models as opposed to $>$90\% for LHFI-A). This indicated that the extra data from ITI metrics weakened the overall relationship between health and covariates.

One possible explanation for this phenomenon is that the LHFI construct was appropriate for describing health using AMBI metrics and the available covariates, but ITI metrics have weak ecological relevance to Richibucto. This is plausible from a qualitative perspective, in light of our prior beliefs as stated in the start of Section \ref{sec-influence}. To quantitatively address this, one may determine additional covariates that can be more appropriately paired with ITI metrics, then model these alongside the original covariates. However, this would require further field activities, and we do not pursue it in this article due to cost constraints.

On the other hand, it is also possible that a) the ITI data were too noisy for Wu's specific LHFI models to detect any patterns in their relationship with the covariates, or that b) the data were not too noisy, but these LHFI models were inadequate for revealing a clear relationship among latent health and covariates.
Scenario a) is certainly conceivable given the type of study at hand, in which data often involve substantial measurement error. If this were the case, there is unlikely much room for the proposed models to be improved upon within the same modelling framework, given that they are already quite ecologically informative.
This leaves us with b) to consider; indeed, Wu's models were merely preliminary models under the general LHFI construct. This could also explain the ``trumping'' phenomenon in the simpler LHFI models. Specifically, distance likely contained much less measurement error than the other covariates, it being easier to measure with precision than the environmental covariates which are intrinsincally more variable in nature. With an LHFI model that was perhaps too simplistic, an effect on health from distance could therefore manifest itself more clearly than effects from other covariates even if all of them are equally important qualitatively.

Existing data can be used immediately to address b), but they require an improved quantitative framework. In light of our prior beliefs, and the fact that the environmental covariates contained specialised information that distance did not, a more sophisticated LHFI modelling framework yet may prove to be helpful in providing a common thread through health, DD and the other ecologically relevant covariates.

Thus, in the rest of this article, we focus on addressing the ``trumping'' phenonmenon in the context of b). To do so, we wish to determine if implementing either or both of the following will allow us to properly identify the nature of the relationship among latent health and the available covariates:
\be
\item Introduce a covariance structure for the metric effects, instead of independence which was assumed for the preliminary models to reduce computational burden.
\item Introduce additional level(s) to the regression hierarchy based upon the known associations between the available covariates.
\ee 
Since these steps pertain to different parts of the LHFI model, below we treat each as a stand-alone investigation. Moreover, we consider LHFI-A models only: introducing extra model complexity for LHFI-A-I models can be impractical for proper inference (e.g.~via MCMC), given that AMBI and ITI metrics are dependent according to their definitions, so that extra model parameters are required to account for this. (For example, see \cite{wu} for the extra parameters involved even when such dependence was only informally accounted for.)

\section{Modelling Covariance of Metric Effects for LHFI-A}\label{sec-cov}

Without additional data, estimating an unstructured covariance matrix may be impractical due to the considerable increase in new parameters in the model. Instead, assuming a structured covariance matrix would be preferred. The form of the LHFI model, for instance, might inherently imply a certain covariance structure. Thus, we examine what covariance structure might be appropriate for the metric effects under previous LHFI-A models, and would then implement them to determine if the extra model complexity can bring about more detectable relationships between latent health and covariates.

\subsection{Extending the LHFI-A Models for Richibucto}

We follow the notation of \citet{wu}. AMBI metrics, denoted by $Y$ in the LHFI framework, are abundances of five disjoint taxonomic groups. Due to the different preconceived directions of their association with health (Table \ref{table-metrics}), we split the metrics into two groups: $s$=''$-$'' for Metrics 3--5 (negatively related to health), and $s$=''+'' for the remaining metrics. In the LHFI model, each member of Group $s$ is modelled as a multinomial random variable. The link function for the GLMM is a generalised logit for $s$=''+'', and an inverted generalised logit for $s$=''$-$'', so that large metric values for $s$=''+'' and ``$-$'' reflect good/neutral and poor health, respectively.

More precisely, let $Y_{i\times j(s)\times k}$, written $Y_{ijks}$ for simplicity, denote the value of the $j$th metric (nested within the $s$th metric group) for the $k$th replicate grab sample at the $i$th site. 
Let $N_{ik}$ denote the \textit{cardinality} (total number of benthic organisms) of the $k$th replicate sample at the $i$th site; and $p_{ijs}$ denote the unknown probability of an organism from the $i$th site belonging to the $j(s)$th taxonomic group. Thus, we have multinomial distributions
\begin{gather}
\begin{split}\label{multi+}
\left\{Y_{i1k+},Y_{i2k+},N_{ik}-Y_{i1k+}-Y_{i2k+}\right\}\ |\ N_{ik},p_{i1+},p_{i2+} \\
\sim\text{Multinomial}(N_{ik};p_{i1+},p_{i2+},1-p_{i1+}-p_{i2+})\,,
\end{split}\\
\begin{split}\label{multi-}
\left\{Y_{i3k-},Y_{i4k-},Y_{i5k-},N_{ik}-Y_{i3k-}-Y_{i4k-}-Y_{i5k-}\right\}\ |\ N_{ik},p_{i3-},p_{i4-},p_{i5-} \\
\sim\text{Multinomial}(N_{ik};p_{i3-},p_{i4-},p_{i5-},1-p_{i3-}-p_{i4-}-p_{i5-})\,.
\end{split}
\end{gather}

Next, let $H_i$ denote the latent health of the $i$th site; and $\theta_s$ and $\beta_{j(s)}$ respectively denote the metric group effect and individual metric effect (both unknown) in the regression model. Then, the linear predictor in the LHFI framework is
\begin{align}
\nu_{ij+} &= \log\frac{p_{ij+}}{1-p_{i1+}-p_{i2+}}\,, \quad  j=1,2, \label{link+} \\
\nu_{ij-} &= \log\frac{1-p_{i3-}-p_{i4-}-p_{i5-}}{p_{ij-}}\,, \quad j=3, 4, 5, \label{link-} \\
\nu_{ijs} &= H_i + \theta_{s} + \beta_{j(s)}\,, \quad s=+,-.  \label{lin.pred} 
\end{align}
For (\ref{lin.pred}), we model $\theta_s$ as a fixed effect and take $\theta_+$=0 (as is customary when considering one of the categories as baseline) to ensure model identifiability, and we model site health and metric effects as random. Specifically, the latent regression of $H_i$ is
\begin{gather}
H_i = \alpha_0+\bs{\alpha}'\bs{x}_i+\eps_i\,,\label{latregr} \\
\eps_i|\sigma_H^2\iid\text{Normal}(0,\sigma_H^2) 
\end{gather}
where $\bs{x}_i$ is the vector of a given combination of the aforementioned covariates,\footnote{In practice, covariate transformation might be necessary to satisfy the linearity of (\ref{latregr}). Covariates (possibly transformed) are then centred to reduce dependence among the $\alpha$s. For a given covariate that is not an interaction, centred data are produced by subtracting from the raw covariate data a constant that is (approximately) equal to the observed covariate mean (averaged over $i$). The ``centred interaction'' between two covariates is taken to be the product of two centred covariates.\label{centred}} $\alpha_0$ and $\bs{\alpha}$ are the unknown coefficients of the corresponding latent regression, and $\eps_i$ is the normally distributed regression error with unknown variance $\sigma_H^2$. 

Note that there is overlap and thus dependency between the two multinomials of (\ref{multi+}) and (\ref{multi-}). \citet{wu} explained that this dependency is crudely accounted for by $\theta_s$; similarly, the mean-zero $\beta_{j(s)}$ crudely accounts for the dependency among $\nu$s within group $s$. Thus, independent $\beta_{j(s)}$s were assumed. More rigorously, we now replace independence of metric effects by
\begin{equation} \label{new.beta}
\bs{\beta}|\bs{\Sigma} \sim \text{MVN}({\bf 0},\bs{\Sigma})\,, \quad
\bs{\beta}=\left[
\begin{matrix}
\bs{\beta}_+ \\
\bs{\beta}_-
\end{matrix}
\right]\,, \quad
\bs{\Sigma} = \left[
\begin{matrix} 
\bs{\Sigma}_{+} & \bs{\Sigma}_{\pm} \\ 
\bs{\Sigma}_{\pm}' & \bs{\Sigma}_{-} 
\end{matrix} \right]
\end{equation}
where $\bs{\beta_+}\equiv  [\beta_{1(+)},\beta_{2(+)}]'$, $\bs{\beta_-}\equiv [\beta_{3(-)},\beta_{4(-)},\beta_{5(-)}]'$, $\bs{\Sigma}$ is the unknown covariance matrix for $\bs{\beta}$, and ``MVN'' denotes the multivariate normal distribution. Thus, $\bs{\Sigma}_{+}$ (2$\times$2), $\bs{\Sigma}_{-}$ (3$\times$3), and $\bs{\Sigma}_{\pm}$ (2$\times$3) denote the covariance matrices for metric groups positively and negatively related to health, and their cross-covariance matrix, respectively.

We use relatively diffuse\footnote{Diffuseness of priors reflects the fact that in the absence of data, we have no clear perception of the properties of the corresponding unknown quantities. In general, diffuseness reduces the need for justification of prior distributional assumptions.} distributions (with the same parametrisation as \citet{wu}) as priors for $\alpha_0,\bs{\alpha},\theta_-$ (univariate Gaussian with mean 0 and variance 100) and $\sigma_H^2$ (inverse-Gamma with unit shape and scale). To complete the Bayesian modelling hierarchy, we must specify the structure for $\bs{\Sigma}$, as we now explore.

\subsection{Dependence Structures of $\beta$s and $\nu$s}\label{sec-dep}
The most general form of $\bs{\Sigma}$ is to take it as fully unstructured, and thus it can be assumed to have an inverse-Wishart (IW) distribution
\begin{gather}\label{unstructured}
\bs{\Sigma}\sim\text{IW}_5
\end{gather}
where ``IW$_d$'' denotes the inverse of a $d\times d$ Wishart matrix with $d$ degrees of freedom and scale matrix equal to the identity, which is a relatively diffuse prior for a $d\times d$ unstructured covariance matrix. Then, one can take advantage of existing MCMC software such as OpenBUGS (\url{http://www.openbugs.info}) for straightforward implementation of the LHFI model, although in our experience, non-trivial hierarchical centring is essential to improve MCMC mixing \citep[see][for details]{chiu.paper,wu}.

However, one concern for assuming (\ref{unstructured}) is that with a small dataset from 18 sites each with only 2 to 3 replicate grab samples, an unstructured $\bs{\Sigma}$ may be only weakly identifiable.\footnote{See \citet{textbook} for a discussion on lack of identifiability.} This issue was encountered by \citet{chiu.paper} for a freshwater benthic dataset that also involved 18 sites with 3 replicates per site, although there were nine metrics altogether. To address this concern, we instead consider a structure for $\bs{\Sigma}$ which involves fewer unknown parameters.

To this end, let us momentarily consider a frequentist's viewpoint.

Without loss of generality, we may drop the subscript $s$ from $p$ and $\nu$ in (\ref{link+})--(\ref{lin.pred}) since the value of $j$ determines $s$. Then, given site $i$, let $\bs{\Sigma}^\nu$ denote the 5$\times$5 covariance matrix whose $(j,j')$th element is $\Sigma_{jj'}^\nu$=Cov($\nu_{ij},\nu_{ij'}$) which does not depend on $i$. In the frequentist context, one can show that $\bs{\beta}$ has covariance matrix
\begin{equation}\label{sig.b}
\bs{\Sigma} =\bs{\Sigma}^\nu-\sigma_H^2\bs{J}_{55}
\end{equation}
where $\bs{J}_{dd'}$ is a $d\times d'$ matrix of 1s. Furthermore, we partition $\bs{\Sigma}^\nu$ into ``+'', ``$-$'' and ``$\pm$'' blocks accordingly.

Thus, as an alternative to (\ref{unstructured}), one may wish to consider the structure of $\bs{\Sigma}^\nu$ while limiting the complexity of the resulting structure ultimately assumed for $\bs{\Sigma}$. For this, we assume that the dependence between the vectors of $\nu_+$s and $\nu_-$s is adequately addressed by $\theta_s$ in (\ref{lin.pred}), and consequently $\bs{\Sigma}_{\pm}^\nu$ is a matrix of 0s. However, we allow $\bs{\Sigma}^\nu_+$ and $\bs{\Sigma}^\nu_-$ to be unstructured.

The form of (\ref{sig.b}) in light of the above consideration suggests that a reasonable structure for $\bs{\Sigma}$ may be
\begin{gather}
\bs{\Sigma}_+|\varsigma\sim\text{IW}_2+\varsigma\bs{J}_{22}\,, \quad
\bs{\Sigma}_-|\varsigma\sim\text{IW}_3+\varsigma\bs{J}_{33}\,, \quad \bs{\Sigma}_{\pm}|\varsigma\equiv\varsigma\bs{J}_{23}\,, \label{IW}\\
\varsigma\sim\text{N}(0,100) \text{ subject to $\bs{\Sigma}$ being positive definite.} \label{vsig}
\end{gather}
Our choice of distributions and hyperparameters in (\ref{IW}) and (\ref{vsig}) yields relatively diffuse priors; it also allows $\bs{\Sigma}$ to be decomposed as the sum of the constant matrix $\varsigma\bs{J}_{55}$ and a random block-diagonal matrix whose blocks are unstructured, thus giving $\bs{\Sigma}$ a general form that mimics that in (\ref{sig.b}) while keeping $\bs{\Sigma}$ positive definite.

Altogether, our extended model comprises (\ref{link+})--(\ref{new.beta}) and (\ref{IW})--(\ref{vsig}). Note that the structure of (\ref{IW})--(\ref{vsig}) for $\bs{\Sigma}$ corresponds purely to (\ref{new.beta}); it will not be the covariance structure in the posterior inference for $\bs{\beta}$.

\subsection{Results of Implementation}\label{impl}

\begin{figure}[t]
\begin{center}
\includegraphics[width=5.2in]{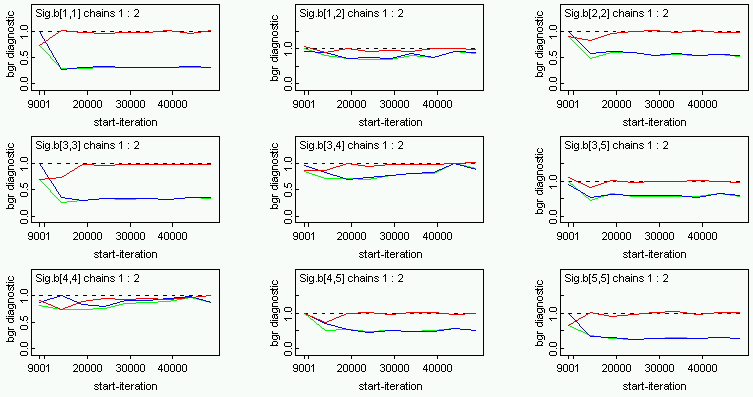}
\end{center}
\caption{Brooks-Gelman-Rubin diagnostics \citep{bugshelp} for the MCMC samples of each matrix element of a block diagonal $\bs{\Sigma}$. ``{\tt Sig.b[i,j]}'' denotes $\Sigma_{ij}$, i.e.~the $(i,j)$th element of $\bs{\Sigma}$. Convergence is suggested by a red curve approaching 1, and green and blue curves approaching the same constant. \label{fig-Sig}}
\end{figure}

\begin{figure}[h!]
\centering
\includegraphics[width=2.8in]{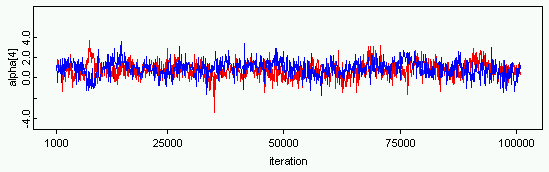}
\includegraphics[width=2.7in]{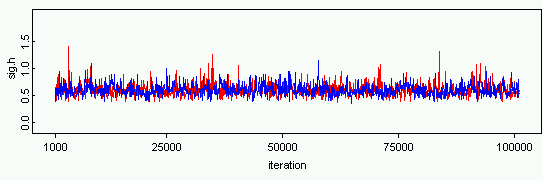}

\includegraphics[width=2.8in]{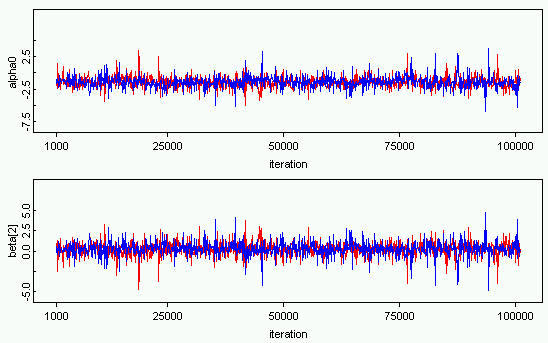}
\includegraphics[width=2.8in]{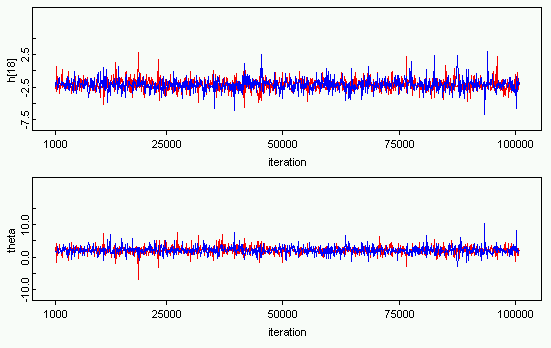}
\caption[]{Trace plots of MCMC iterations (thinned by 100) from the posterior for fitting (\ref{link+})--(\ref{unstructured}); each covariate in (\ref{latregr}) has been centred (see Footnote \ref{centred}).
Counter-clockwise from top-left: regression coefficient of the centred 
interaction $\text{log(depth)}\times\text{log(SC)}$, intercept $\alpha_0$, random effect $\beta_2$, fixed effect $\theta_-$, latent health $H_{18}$, and standard deviation $\sigma_H$. Trace plots for all other non-$\bs{\Sigma}$ model parameters show similar patterns that suggest convergence after a burn-in of merely 1,000.\label{fig-non-Sig}}
\end{figure}

For the investigation, we focus on a single LHFI-A model involving the covariates DD, log(depth), log(SC) and log(depth)$\times$log(SC).  Bayesian parameter estimates and credible intervals then are compared to the LHFI-A models from \citet{wu} with the same set of covariates. 
Though, instead of fitting (\ref{IW})--(\ref{vsig}) which would require an implementation outside of OpenBUGS, we implemented a block diagonal $\bs{\Sigma}$ (i.e.~$\varsigma$$\equiv$0) as a limiting case. 
Inference for Richibucto latent health as a whole\footnote{``Health inference as a whole'' here refers to the ranking of sites according to the posterior distributions of $H_i$s.} from this limiting case was very similar to that from a diagonal $\bs{\Sigma}$ as assumed by \citet{wu}, suggesting reasonable robustness of the health inference to $\bs{\Sigma}$. Additionally, the concern of weak identifiability associated with a non-diagonal $\bs{\Sigma}$ proves to be somewhat irrelevant for these data, as our two independently generated MCMC chains mixed readily after a burn-in of around 20,000 iterations: parameters of the non-diagonal $\bs{\Sigma}$ required this longer burn-in (Fig.~\ref{fig-Sig}), although all other model parameters each required a burn-in of only 1,000\footnote{For a given model, inference for model parameters as a whole was always based on the longest burn-in required.} (Fig.~\ref{fig-non-Sig}). 

Finally, we observe that the significance of the covariates was also virtually unaffected by assuming a $\bs{\Sigma}$ that is more complex than that for \citet{wu}.

\section{An Additional Level in the Latent Regression}

The investigation of Section \ref{sec-cov} suggested that the extra complexity from a non-trivial dependence structure among metric effects does not help to clarify the relationships between latent health and covariates given the current Richibuto data. In this section, we revert to the naive independence assumption for $\bs{\beta}$ but introduce extra model complexity through an additional level in the regression of latent health on covariates.

\begin{figure}[h!]
\centering
\fbox{\includegraphics[width=8cm]{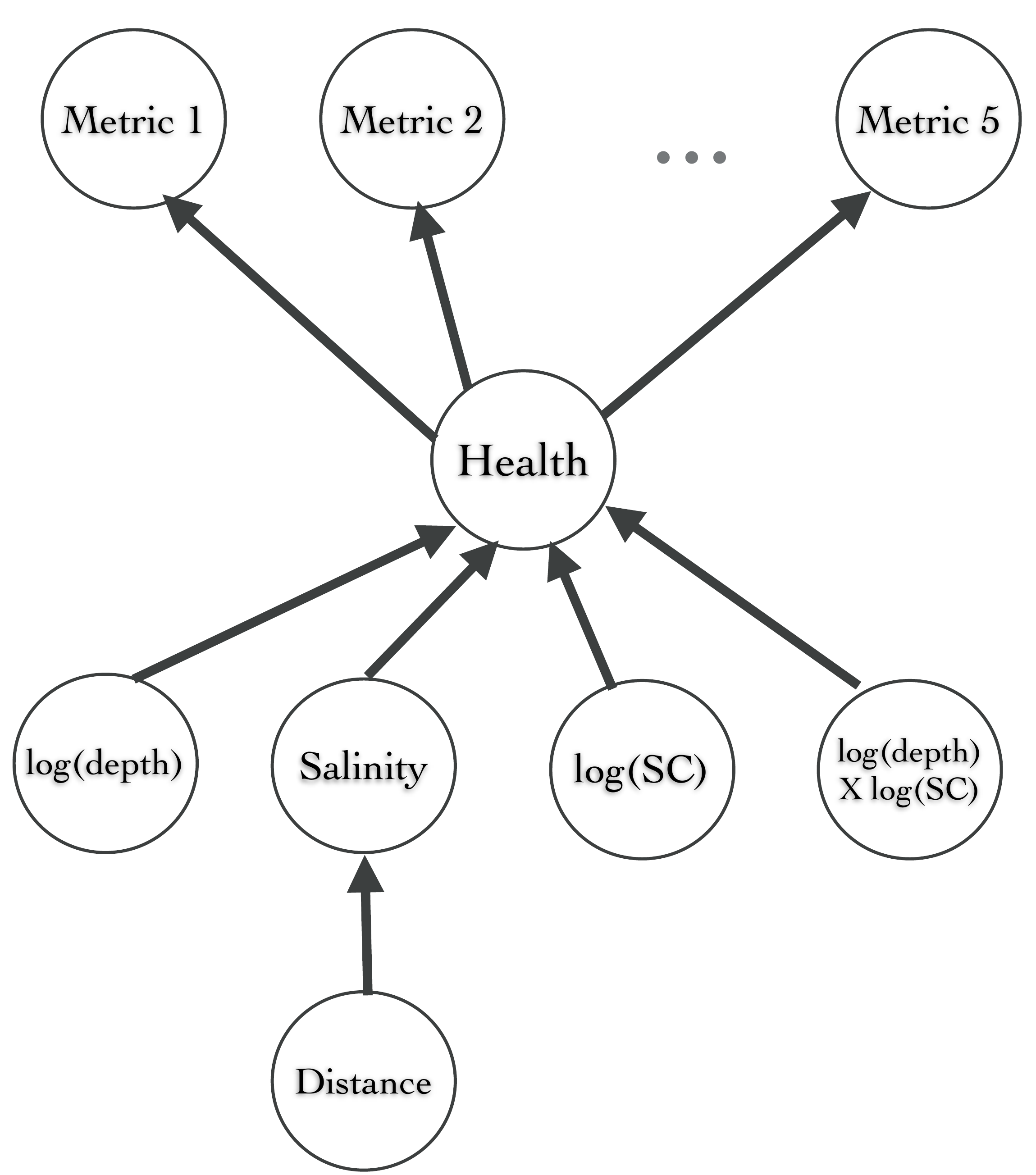}}
\caption{Graphical depiction of regressing salinity on DD as an additional level in the hierarchical latent health model, while, as usual, health is the response of salinity and other covariates, and AMBI metrics are responses of latent health.}\label{schem_4cov}
\end{figure}

Specifically, although the strong correlation between salinity and DD reflects ecological reasoning for coastal sea waters entering an estuary, it is the only clear empirical relationship detected among the available covariates. Therefore, instead of considering salinity and DD to be complementary covariates, we now take salinity as a response of DD, and in turn, latent health as a response of salinity and the remaining covariates from Section~\ref{impl} (Fig.~\ref{schem_4cov}).

Then, model statements of this section include (\ref{multi+})--(\ref{new.beta}), and additionally,
\begin{gather}
x_{\text{sal,}i} = \alpha_\text{DD} x_{\text{DD,}i}+\delta_i\,,
 \label{new.regr} \\
\delta_i|\sigma_{\delta}^2 \ \iid \text{Normal}(0,\sigma_{\delta}^2) \label{deltadist}
\end{gather}
where $\bs{\Sigma}=\sigma^2_\beta\mathbb{I}$ in (\ref{new.beta}), $\mathbb{I}$ is the identity matrix, and $\bs{x}_i$ in (\ref{latregr}) denotes the vector of centred covariates for site $i$ including salinity $x_{\text{sal,}i}$ (and possibly other covariates) but excluding DD $x_{\text{DD},i}$. Hence, (\ref{latregr}) and (\ref{new.regr}) can be collapsed as follows:
\begin{equation}\label{saldistcombined}
H_i = \alpha_0+ \bs{\alpha}_{-\text{sal}}'\bs{x}_{-\text{sal},i}+\alpha_\text{sal}\alpha_\text{DD}x_{\text{DD},i}+\alpha_\text{sal}\delta_i+\eps_i 
\end{equation}
where $\bs{\alpha}_{-\text{sal}}$ is $\bs{\alpha}$ with $\alpha_\text{sal}$ removed, and similarly for $\bs{x}_{-\text{sal},i}$. Thus, (\ref{saldistcombined}) regards salinity as an implicit covariate, so that when latent health is explicitly regressed on $\bs{x}_{-\text{sal},i}$ and DD, the implicit covariate decomposes the total error variation into
Var($\alpha_\text{sal}\delta_i+\eps_i|\alpha_\text{sal},\sigma_\delta^2,\sigma_H^2$) = $\alpha_\text{sal}^2\sigma_\delta^2+\sigma_H^2$. Hence, a smaller ratio $\sigma_H^2/(\alpha_\text{sal}^2\sigma_\delta^2+\sigma_H^2)$ reflects a higher contribution from the implicit covariate towards explaining the total error variation of the latent health regression.

We employ the same inverse-Gamma prior for $\sigma_H^2$ as well as $\sigma_\delta^2$. Univariate N(0, 100) priors are employed for $\theta_-,\alpha_0$, and vector components of $\bs{\alpha}$, with one exception: Cor($\alpha_\text{sal},\alpha_\text{DD}$) = $\rho\ne$ 0 is additionally considered, where $\rho\sim$ Unif($-$1, 1) {\it a priori}.

\subsection{Results}
\begin{figure}[h!]
\centering
\includegraphics[width=14cm]{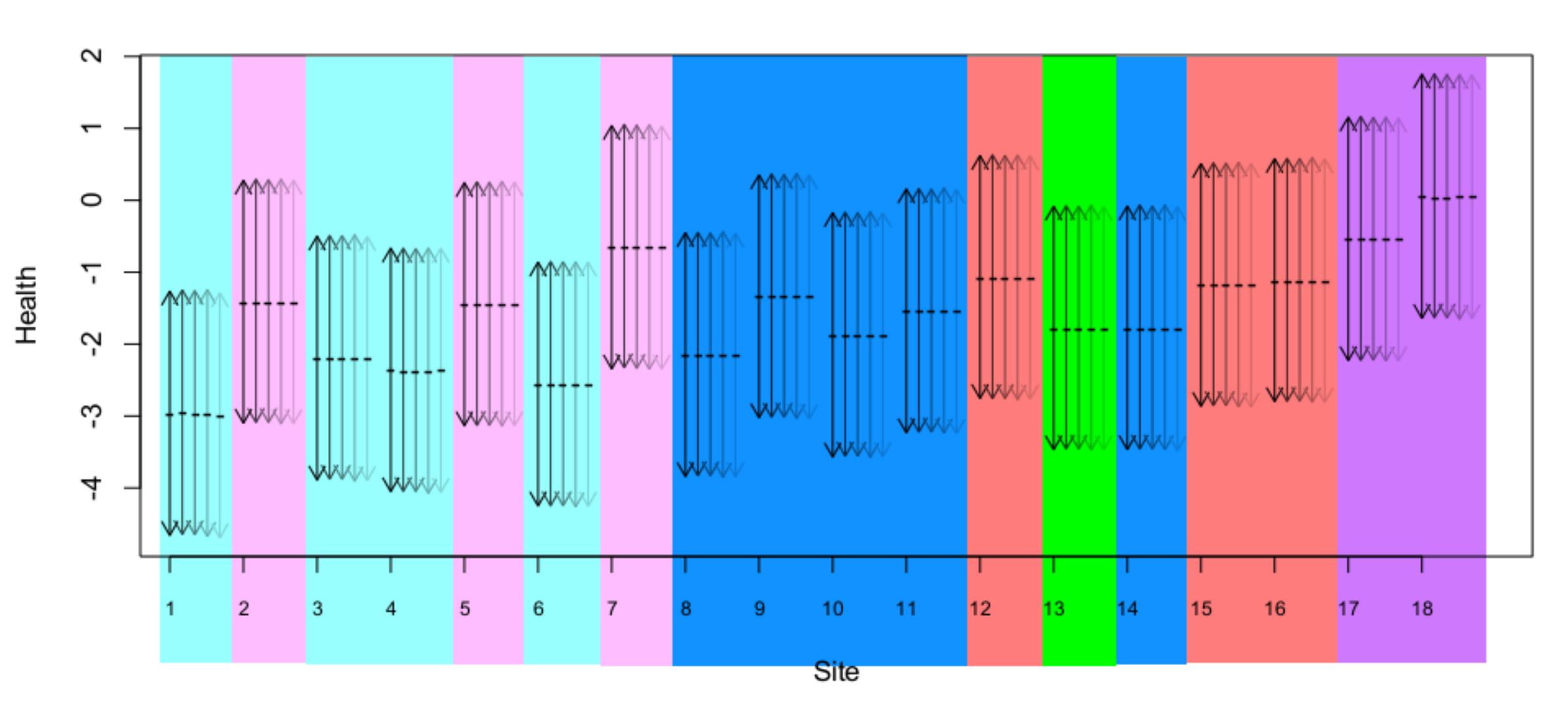}
\caption{LHFI scores (posterior means marked by `-') and 95\% CIs of site health (arrows from dark to light), based on Models (4), (5), (1), (2), and (3) of Table \ref{stats}, respectively. \cite{luke} partition Richibucto sites into six groups according to their benthic community composition: red (lower channel), violet (estuarine mouth), green, blue (lower shallow), turqoise (upper shallow), and pink (upper channel).}
\label{healthplot}
\end{figure}

\begin{figure}[h!]
\centering
\mbox{ 
\subfigure[]
{    \label{plot.alpha} \includegraphics[width=7cm]{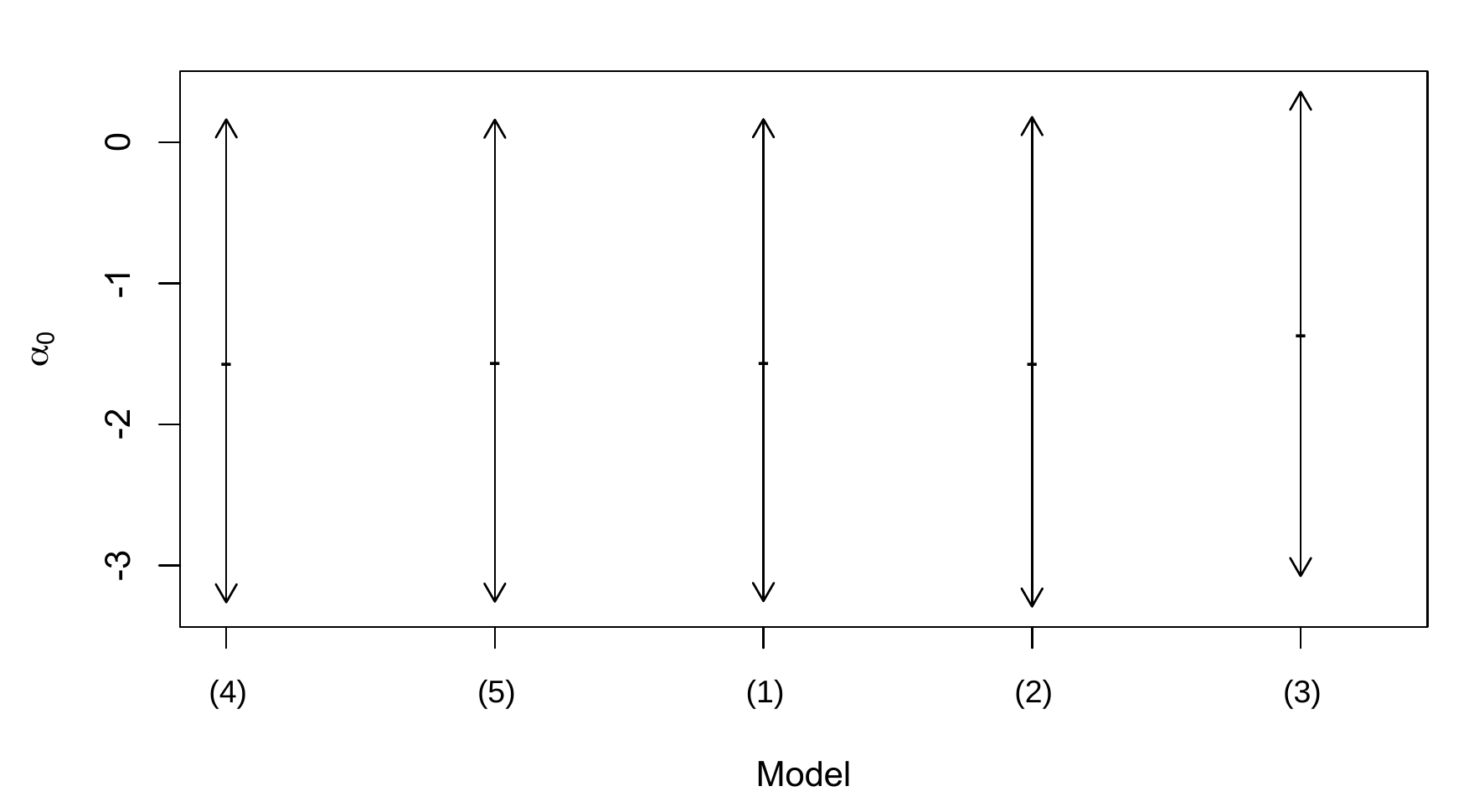}
} \quad
\subfigure[]
{    \label{plot.theta} \includegraphics[width=7cm]{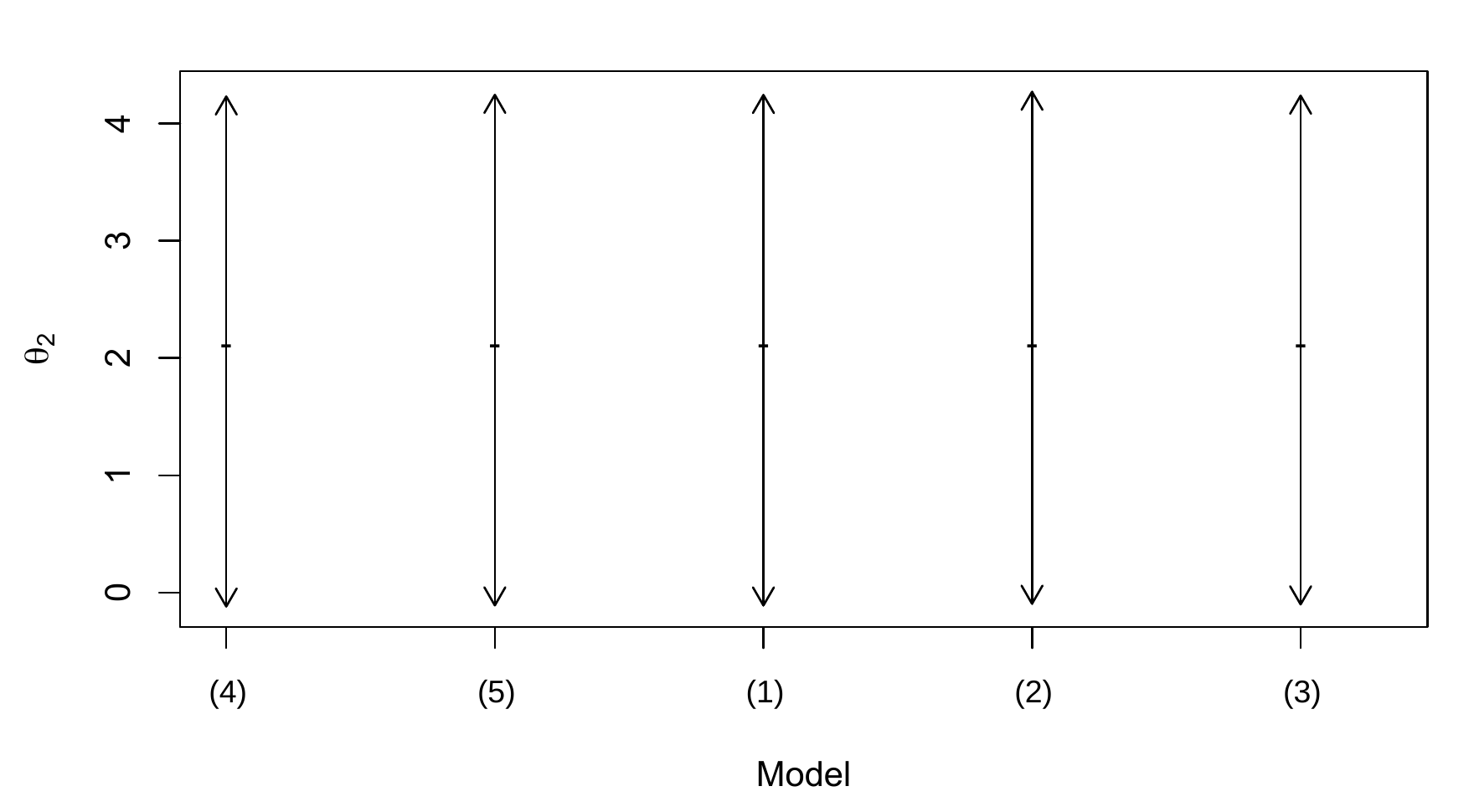}
}}
\mbox{
\subfigure[]
{    \label{plot.beta12} \includegraphics[width=7cm]{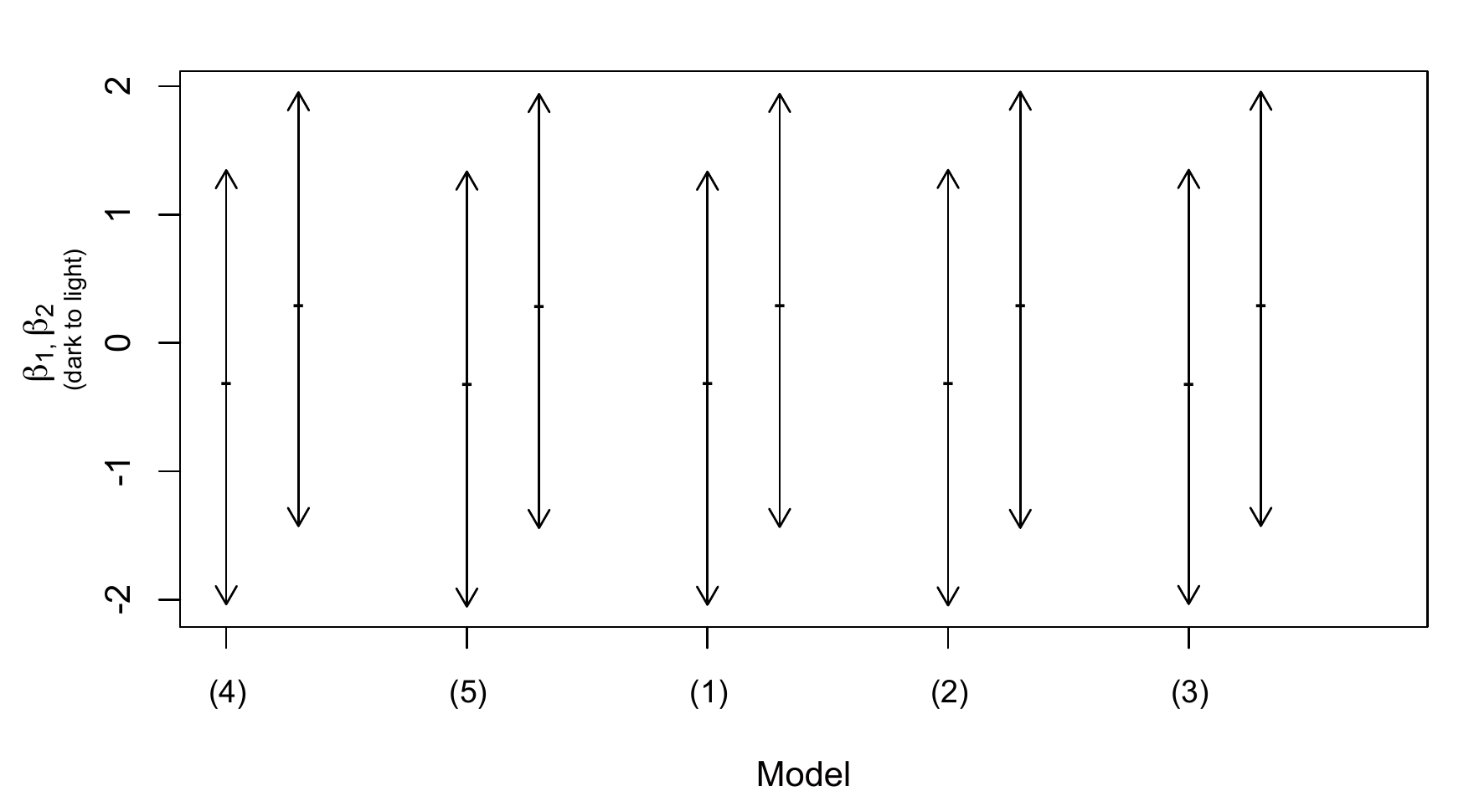}
} \quad
\subfigure[]
{    \label{plot.beta345} \includegraphics[width=7cm]{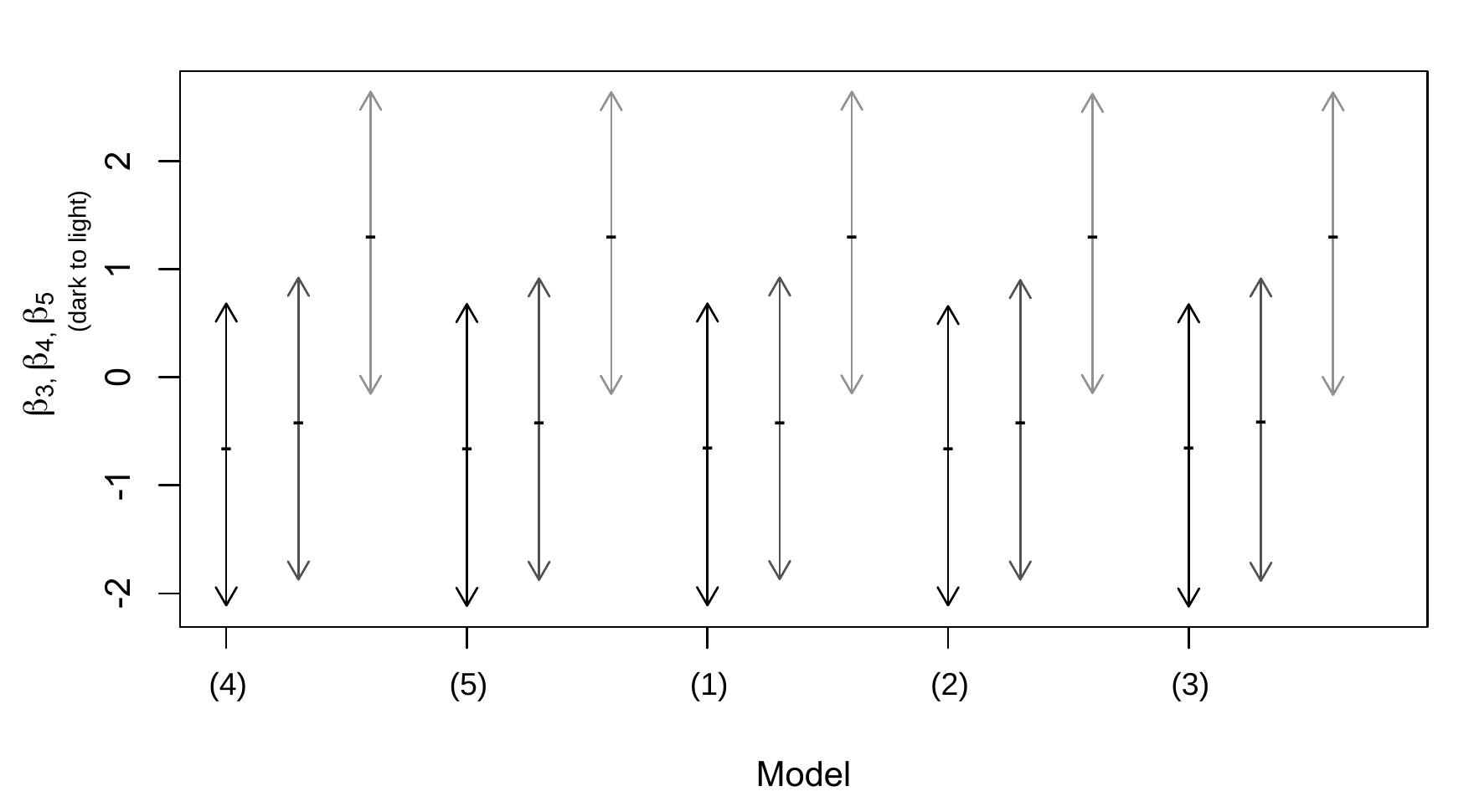}
}}
\mbox{\subfigure[]
{    \label{plot.sigh} \includegraphics[width=7cm]{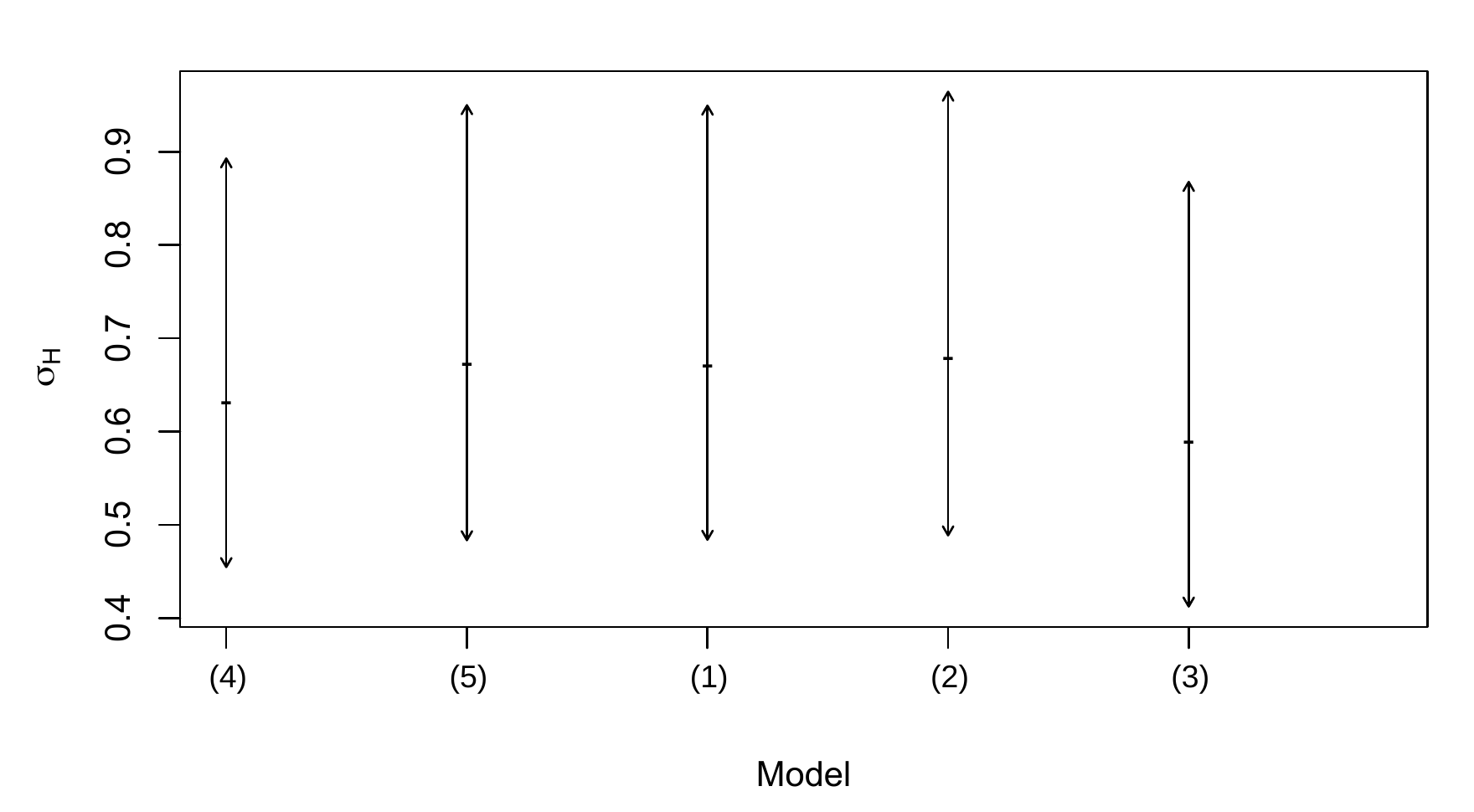}
}\quad
\subfigure[]
{    \label{plot.sigb} \includegraphics[width=7cm]{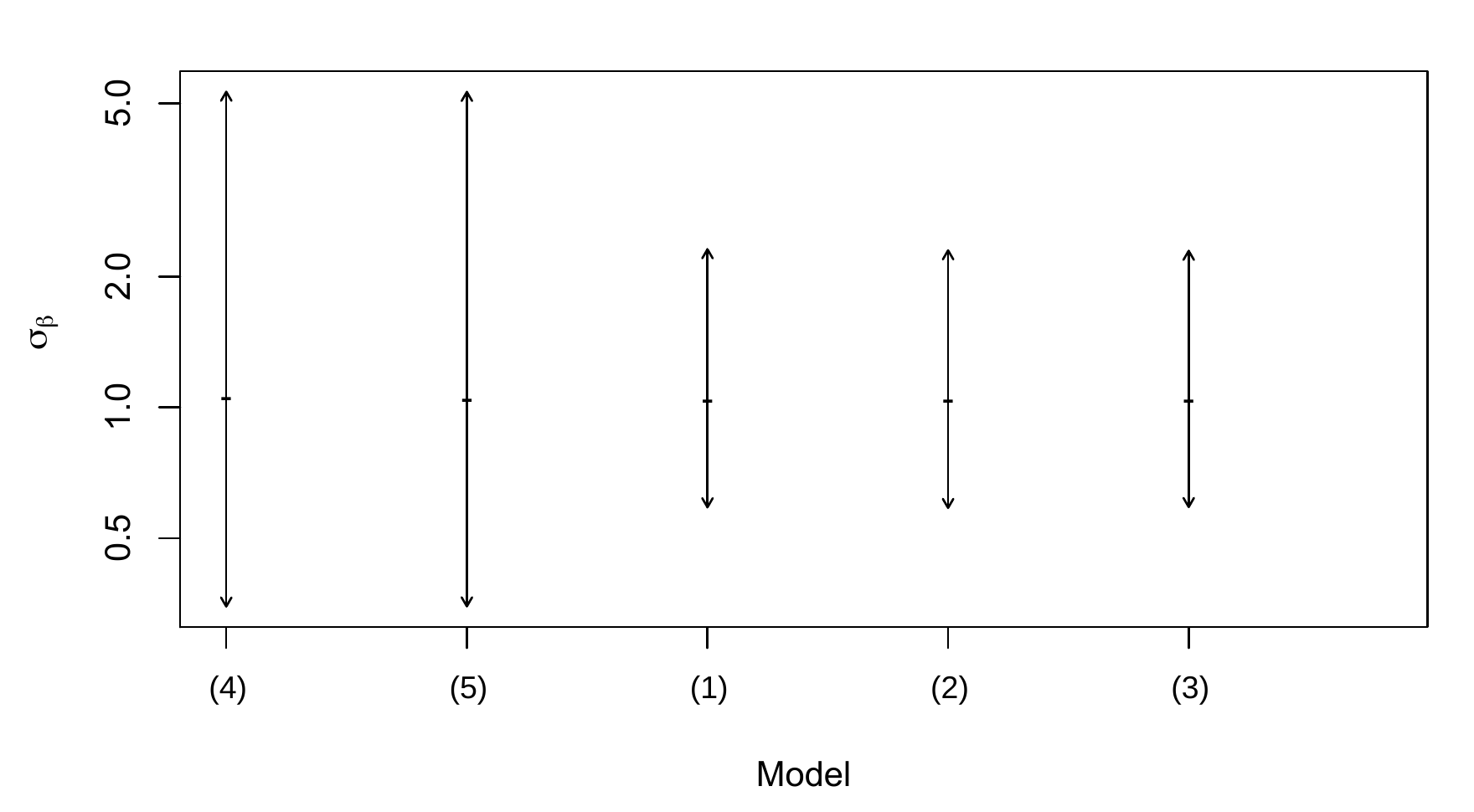}
}}
\caption{Posterior means and 95\% CIs for (a) $\alpha_0$, (b) $\theta_2$, (c) elements of $\bs{\beta}_+$, (d) elements of $\bs{\beta}_-$, (e) $\sigma_{H}$, and (f) $\sigma_{\beta}$ (plotted on log-scale).}
\label{plot.other} 
\end{figure}

Inference summaries appear in Table~\ref{stats}, in which Models (1)--(3) each comprises two levels of covariates (expressions (\ref{multi+})--(\ref{new.beta}) and (\ref{new.regr})--(\ref{deltadist})), and Models (4) and (5) --- provided as a comparison --- each comprises a single level of covariates (expressions (\ref{multi+})--(\ref{new.beta}) only). Posterior means for latent health along with their 95\% posterior credible intervals (CIs) appear Fig.~\ref{healthplot}; those for $\alpha_0$, $\beta_{j(s)}$, $\theta_s$, $\sigma_{H}$, and $\sigma_{\beta}$ appear in Fig.~\ref{plot.other}.
\begin{table}[h!]
\centering
\caption{Selected summary statistics of posterior draws. Boldfaced CI limits suggest that the corresponding parameter differs from 0 with high credibility.}
\begin{tabular}{cccrrrrcc}\\\hline
&&&&& \multicolumn{2}{c}{95\% CI} \\ \cline{6-7}
& Model &  & Mean & Median & 2.5\% & 97.5\% & DIC\\ \hline\\
(1) & {\it sal}-on-{\it DD} only; & $\alpha_0$ & $-$1.56 & $-$1.57 & $-$3.25 & 0.16 & 4417 \\ 
& Cor($\alpha_\text{sal},\alpha_\text{DD}$)=0 & $\alpha_\text{sal}$ & 0.39 & 0.39 & {\bf 0.13} & {\bf 0.64} \\ 
& & $\alpha_\text{DD}$ & 0.77 & 0.77 & {\bf 0.54} & {\bf 1.00} \\ 
& & $\sigma_{\beta}$ & 1.12 & 1.01 & 0.59 & 2.31 \\ 
& & $\sigma_{\delta}$ & 0.70 & 0.68 & 0.51 & 0.98 \\ 
& & $\sigma_{H}$ & 0.67 & 0.65 & 0.48 & 0.95\\ 
& & $\theta_2$ & 2.09 & 2.10 & $-$0.11 & 4.24 \\
& & $\frac{\sigma_H^2}{\alpha_\text{sal}^2\sigma_\delta^2+\sigma_H^2}$ & 0.54 & 0.55 & 0.31 & 0.76 \\ \\
(2) & {\it sal}-on-{\it DD} only;  & $\alpha_0$ & $-$1.57 & $-$1.58 & $-$3.29 & 0.18 & 4419 \\ 
& Cor$(\alpha_\text{sal},\alpha_\text{DD})$=$\rho$ & $\alpha_\text{sal}$ & 0.49 & 0.49 & {\bf 0.28} & {\bf 0.71}  \\ 
& & $\alpha_\text{DD}$ & 0.59 & 0.59 & {\bf 0.35} & {\bf 0.79}  \\ 
& & $\rho$ & $-$0.94 & $-$0.96 & {\bf $-$1.00} & {\bf $-$0.78} \\ 
& & $\sigma_{\beta}$ & 1.12 & 1.01 & 0.59 & 2.30 \\ 
& & $\sigma_{\delta}$ & 0.74 & 0.72 & 0.53 & 1.07 \\ 
& & $\sigma_{H}$ & 0.68 & 0.66 & 0.49 & 0.96 \\ 
& & $\theta_2$ & 2.10 & 2.10 & $-$0.10 & 4.27 \\ 
& & $\frac{\sigma_H^2}{\alpha_\text{sal}^2\sigma_\delta^2+\sigma_H^2}$ & 0.59 & 0.59 & 0.37 & 0.79 \\ \\
(3) & {\it log(depth), log(SC)}, & $\alpha_0$ & $-$1.37 & $-$1.37 & $-$3.08 & 0.36 & 4417 \\ 
& {\it log(depth)$\times$log(SC)}, & $\alpha_\text{dep}$ & 0.16 & 0.16 & $-$0.63 & 0.94 \\ 
& and {\it sal}-on-{\it DD}; & $\alpha_\text{sal}$ & 0.42 & 0.42 & {\bf 0.17} & {\bf 0.67} \\ 
& Cor($\bs{\alpha}$)={\bf 0} & $\alpha_\text{SC}$ & $-$0.83 & $-$0.83 & $-$1.74 & 0.08 \\ 
& & $\alpha_\text{DD}$ & 0.77 & 0.77 & {\bf 0.54} & {\bf 1.00} \\ 
& & $\alpha_\text{dep$\times$SC}$ & 1.88 & 1.88 & {\bf 0.28} & {\bf 3.49} \\ 
& & $\sigma_{\beta}$ & 1.12 & 1.01 & 0.59 & 2.29 \\ 
& & $\sigma_{\delta}$ & 0.70 & 0.68 & 0.51 & 0.98 \\ 
& & $\sigma_{H}$ & 0.59 & 0.57 & 0.41 & 0.87 \\ 
& & $\theta_2$ & 2.09 & 2.10 & $-$0.10 & 4.24\\
& & $\frac{\sigma_H^2}{\alpha_\text{sal}^2\sigma_\delta^2+\sigma_H^2}$ & 0.63 & 0.63 & 0.37 & 0.87 \\ \\
(4) & {\it DD} only & $\alpha_0$ & $-$1.57 & $-$1.58 & $-$3.26 & 0.16 & 4380 \\ 
& & $\alpha_\text{DD}$ & 0.37 & 0.37 & {\bf 0.16} & {\bf 0.58} \\ 
& & $\sigma_{\beta}$ & 1.48 & 1.03 & 0.35 & 5.32 \\ 
& & $\sigma_{H}$ & 0.63 & 0.61 & 0.45 & 0.89 \\ 
& & $\theta_2$ & 2.09 & 2.10 & $-$0.12 & 4.23 \\ \\
(5) & {\it sal} only & $\alpha_0$ & $-$1.57 & $-$1.57 & $-$3.26 & 0.16 & 4379 \\ 
& & $\alpha_\text{sal}$ & 0.39 & 0.39 & {\bf 0.13} & {\bf 0.64} \\ 
& & $\sigma_{\beta}$ & 1.48 & 1.02 & 0.35 & 5.32\\ 
& & $\sigma_{H}$ & 0.67 & 0.66 & 0.48 & 0.95  \\ 
& & $\theta_2$ & 2.10 & 2.10 & $-$0.11 & 4.24 \\ \\
\hline
\end{tabular}

\label{stats}
\end{table}

For the assessment and prediction of latent health, the model parameters of main interest include $H_i,\sigma_{H}$, and $\bs{\alpha}$, while the rest of the parameters are regarded as nuisance \citep{wu}. 

It is evident from the 95\% CIs in Table \ref{stats} that, when DD is considered a driver of salinity, the two are simultaneously relevant to explaining latent health. In fact, 0 is excluded from the 99\% CI (not shown) for both $\alpha_\text{sal}$ and $\alpha_{DD}$ in each of Models (1)--(3), suggesting very high credibility for the two-level structure. The CIs from Model (3) suggest that the interaction log(depth)$\times$log(SC) is an additional credible driver of latent health, complementing the explanatory capacity of salinity-on-DD. This is the first time that a scientifically sensible model has been successfully constructed to rigorously identify the collective explanatory capacity of salinity, DD, depth, and SC --- all regarded {\it a priori} as qualitatively important --- towards site health in the Richibucto ecosystem.

For Models (1)--(3), the posterior mean for the ratio $\sigma_H^2/(\alpha_\text{sal}^2\sigma_\delta^2+\sigma_H^2)$ ranges from around 0.55 to 0.65, respectively; corresponding 95\% CIs suggest a smallest ratio for Model (1). Thus, despite the high credibility of the correlation between $\alpha_\text{sal}$ and $\alpha_\text{DD}$ in Model (2) and of the influence on health from (the interaction between) depth and SC in Model (3), the least complex Model (1) provides slightly clearer evidence for the explanatory capacity of the salinity-on-DD structure. In terms of the model's predictive power, the least and most complex among the three models share the same deviance information criterion (DIC) \citep{dic} which is slightly smaller (better) than that of Model (2). This predictive power corresponds to observed AMBI metrics (not latent health) being predicted by the model. To assess the model's predictive ability for latent health, one could conduct a simulation study in which unobservable $H_i$ values are generated then estimated, although such an approach for hierarchical models has its shortcomings \citep{marshall} or requires intensive computations \citep{dey} that may be impractical.

Instead, we compare CIs for $H_i$ among models; in Fig.~\ref{healthplot}, they appear nearly identical across all Models (1)--(5), i.e.~the inference for health is essentially equally credible across various models. Within model, the ranking of sites according to their LHFI scores and associated CIs do not coincide with the clustering by \cite{luke} based on similarity in benthic community composition (highly correlated with site location). This indicates that the LHFI approach does not merely represent community composition or site location; instead, it rigorously and comprehensively models biotic indicators, abiotic drivers, the abstract notion of health, and the relationship among them. Note that the health CIs from the 18 sites mutually overlap, suggesting that the small dataset does not allow us to distinguish sites at a high credible level based on health; this was also the case for those models by \cite{wu}, all with single-level covariates. Despite (i) suboptimal distinguishability and (ii) weaker predictive power for AMBI metrics compared to the single-covariate Models (4)--(5), our two-level-covariate Models (1)--(3) clearly resolved the earlier counterintuitive phenomenon of covariates not being simultaneously significant. Indeed, (i) is an improvement over conventional methods in quantitative rigour due to the integrated manner from which our uncertainty estimates are obtained. Moreover, (ii) is of secondary concern when the response of key interest is $H_i$ instead of the metrics. Technical note: The LHFI framework is built on the fundamental principles of analysis-of-covariance, so that one can only interpret $H_i$ values in a relative sense. However, \cite{chiu.paper} explain that including in the study any site that is qualitatively pre-identified as very healthy or very unhealthy would facilitate the interpretation of the magnitude of $H_i$ for an individual $i$. This is slightly different from the approach of \cite{lopez} who include sites that span the spectrum of individual covariates.

Finally, aside from nearly identifical CIs for $H_i$ across models, Fig.~\ref{plot.other} indicates that the five models perform equally well with respect to the uncertainty (width of CIs) of various nuisance parameters, but with one exception: two-level-covariate models have clearly less uncertainty in their inference for $\sigma_\beta$ (Fig.~\ref{plot.sigb}). As this parameter directly contributes to the uncertainty in the linear predictor $\nu$ of AMBI metrics, Fig.~\ref{plot.sigb} indicates that having two levels of covariates lead to more reliable inference for the model as a whole.

\section{Conclusions}

Unlike conventional multimetric health indices, the integrated LHFI approach yields health scores, asseses the influence of health drivers, and provides their associated uncertainty, all in a single, unified analysis for a given model.

LHFI models by \cite{wu} with single-level covariates were satisfactory as preliminary models for the 18 Richibucto sites, but lacking the important ability to rigorously identify relationships between health and abiotic drivers that are deemed ecologically important for the Richibucto estuarine system. \citet{wu} proposed, without implementation, two ways to address this issue: (a) to introduce a covariance structure on the random metric effects, and (b) to introduce additional layers of regression given preconceived relationships among the covariates. In this paper, we implemented (a) and (b) with AMBI biotic metrics only, but the approach would be applicable in principle to combining AMBI and ITI biotic metrics. Though, with merely 18 sites in Richibucto, combined AMBI-ITI models by \cite{wu} suggest that ITI metrics potentially weaken any signal in the health-covariate relationship.

In this paper, we have found (a) to be ineffective. Various block diagonal covariance structures were considered, the most general of which was reported in this paper. However, none substantially affected the health inference nor improved credible levels of the covariates. 

On the other hand, our efforts on (b) proved to be well rewarded. An additional layer of covariates based upon the empirical relationship between salinity and distance downstream allowed the model to identify the simultaneous significance of distance and those abiotic covariates that were previously shown by \cite{wu} to be significant only when distance was excluded. Moreover, we also show that model inference is more reliable overall when compared to single-level-covariate models. Thus, our two-level-covariate modelling framework more comprehensively exploits the ecological relationship among health, biotic metrics, and abiotic covariates, and it yields less uncertainty in model inference. We implemented three variants of the two-level-covariate model: (a) salinity-on-distance alone, with {\it a priori} independent regression coefficients and metric effects, (b) same as (a) but assuming bivariate regression coefficients, and (c) same as (a) but including channel depth and silt-clay content (both on the log scale), as well as their interaction. Overall, (a)--(c) are almost equal in statistical performance, with slightly better predictive power of biotic metrics by (a) and (c). Finally, (a) corresponds to marginally stronger evidence for the two-level structure between salinity and distance to influence site health.

Although field data, especially biotic data, are costly to collect and process for the use in quantitative assessment of ecosystem health, our work has shedded light on one practical concern: more than 18 sites and/or more precise measurements on abiotic covariates are needed in order for the LHFI framework to rigorously distinguish Richibucto sites according to AMBI and/or ITI metrics as indicators of ecosystem health. This point will be considered as part of future health monitoring and conservation efforts for the Richibucto estuarine system.

\subsection*{Acknowledgments}
\small
The research of this paper was conducted through research assistantships to M.~Wu, funded by an NSERC Discovery Grant to G.~Chiu and an NSERC Discovery Strategic Project Grant to J.~Grant (Dalhousie University) subcontracted to G.~Chiu. Part of this paper was written as an activity in a CSIRO Water for a Health Country Flagship Appropriation Research Project under the Ecosystem Responses to Flow Stream in the Ecosystems and Contaminants Theme. We thank Profs.~J.~Grant and M.~Dowd (Dalhousie University) for their advice and support throughout the duration of this research. G.~Chiu thanks Dr.~A.~Zwart (CSIRO Mathematics, Informatics and Statistics) for his valuable comments.
\normalsize


\renewcommand{\bibname}{References}

\bibliographystyle{apalike}
\addcontentsline{toc}{chapter}{\textbf{References}}
\bibliography{arXiv}

\nocite{*}

\end{document}